\documentclass[twocolumn,showpacs,amssymb,preprintnumbers,prl]{revtex4}
\usepackage{rotating,epsfig,dcolumn}
\usepackage{graphicx,bm}
\begin{document}
\title{Isobaric multiplet yrast energies and isospin non-conserving forces}
\author{ A. P. ~Zuker$^a$, S. M. ~Lenzi$^b$,
  G.~Mart\'{\i}nez-Pinedo$^{c}$, 
  A.~Poves$^d$}
\affiliation{
(a) IReS, B\^at27, IN2P3-CNRS/Universit\'e Louis
Pasteur BP 28, F-67037 Strasbourg Cedex 2, France\\
(b) Dipartimento di Fisica and INFN, Padova, Italy\\
(c) Department f\"ur Physik und Astronomie, Universit\"at Basel, CH-4056
Basel, Switzerland\\
(d) Departamento de Fisica Te\'orica C-XI Universidad Aut\'onoma
de Madrid, E-28049, Madrid, Spain}
\date{\today}
\begin{abstract}
  The isovector and isotensor energy differences between yrast states
  of isobaric multiplets in the lower half of the $pf$ region are
  quantitatively reproduced in a shell model context. The isospin
  non-conserving nuclear interactions are found to be at least as
  important as the Coulomb potential. Their isovector and isotensor
  channels are dominated by $J=2$ and $J=0$ pairing terms,
  respectively. The results are sensitive to the radii of the states,
  whose evolution along the yrast band can be accurately followed.
\end{abstract}
\maketitle 

The electrostatic energy of a sphere of radius $R$ and charge $Ze$ is
easily calculated to be $E_{C}=3e^2Z^2/5R$. It is under this guise
that the Coulomb field enters the Bethe Weiz\"acker mass formula, and
becomes a basic quantity in nuclear structure. Direct evidence of
entirely Coulomb effects has long been available from displacement
energies between mirror ground states (MDE), and in the last decade
from differences in {\em excitation} energies of yrast bands in mirror
nuclei (MED)~\cite{cameron,49,47-49,51,50fe}\footnote{As isospin non conserving
  nuclear forces play an important role we propose to replace the
  traditional acronym for Coulomb displacement energies---CDE---by
  MDE, where M stands for mirror. Similarly CED becomes MED.}.

The MDE energies range from few to tens of MeV. They are given mainly
by $E_{C}$, but precise calculations were found to be unexpectedly
inaccurate---the Nolen-Schiffer anomaly~\cite{NS69}---and 
revealed the necessity to introduce charge symmetry breaking (CSB)
nuclear potentials~(\cite{mac01} and references therein). The
anomaly is now under control {\em to within shell
  effects}~\cite{bro00b,duf02}, which we define as deviations from a
Bethe Weisz\"acker-type formula involving only number of particles
($A$), isospin ($T$), and its third component ($T_z$). 

The  MED are
defined by ($Z_>$ and $Z_<$ are the largest and smallest $Z$ in the
multiplet, and $E_J$ are the yrast excitation energies)
\begin{equation}
  \label{eq:med}
  {\rm MED}_J=E_J(\bar Z+T)-E_J(\bar Z -T),\hspace{.2cm} \bar Z=\frac{Z_>+Z_<}{2}.
\end{equation}
The observed MED are very small (of the order of 10-100 keV), and
entirely due to shell effects. Recently, the experimental information
on yrast bands has been extended to isospin triplets~\cite{50mn,46cr}, thus
determining new quantities, the TED given by
\begin{equation}
  \label{eq:ted}
  {\rm TED}_J=E_J(\bar Z+1)+E_J(\bar Z-1)-2E_J(\bar Z).
\end{equation}  
Both measurements are needed to achieve a clear understanding of the
interplay between the Coulomb potential $V_C$, and $V_B$, the isospin
breaking nuclear interaction.

To analyze them, we start by writing the isovector, $\beta_{\bm
  r}^{(1)}$, and isotensor, $\beta_{\bm r}^{(2)}$, contributions to
$V_B$, as linear combinations of two body matrix elements in
neutron-proton ($\nu\pi$) formalism ($\bf{r} \equiv$ $ r_1\, r_2\,
r_3\, r_4$, where $r_i$ is a subshell):
\begin{equation}
  \label{eq:vb}
\beta_{\bm r}^{(1)}=V^{\pi\pi}_{B\bm{r}}-V^{\nu\nu}_{B\bm{r}},
\hspace{.2cm} 
\beta_{\bm
  r}^{(2)}=V^{\pi\pi}_{B\bm{r}}+V^{\nu\nu}_{B\bm{r}}-2V^{\pi\nu}_{B\bm{r}}. 
\end{equation}

The isoscalar contribution $\beta_{\bm
  r}^{(0)}=V^{\pi\pi}_{B\bm{r}}+V^{\nu\nu}_{B\bm{r}}+V^{\pi\nu}_{B\bm{r}}$
is nil in $V_B$ , while for $V_C$ we have $\beta_{C{\bm
    r}}^{(0)}=\beta_{C{\bm r}}^{(1)}=\beta_{C{\bm
    r}}^{(2)}=V^{\pi\pi}_{C\bm{r}}$.

The MED are entirely of isovector origin and the first exact shell
model calculations in the full $pf$ shell indicated that $V_C^{\rm
  ho}$, i.e., calculated in the harmonic oscillator (ho) basis fails
to give a satisfactory description~\cite{47-49}. The way out proposed
in this reference consisted in replacing the harmonic oscillator
matrix elements $V_{C\bf{f_{7/2}}}^{\rm ho}$ by empirical ones derived
from the $A=42$ spectrum {\em which are very different}. Therefore, it
was hard to attribute the replacement to a renormalization of $V_C$,
expected to be small. But it was equally hard to think in terms of CSB
{\em precisely because the effect was so large}.  Nevertheless the
ansatz (or variants of it) worked quite well, and subsequent
calculations incorporated it~\cite{51,50fe} leading eventually to
(almost) full quantitative agreement~\cite{zuk01} for the MED in
$A=47$, 49, 50 and 51. When the isotensor TED data came in, it became
clear that {\em both} charge independence breaking~\cite{mac01b} and CSB
had to be invoked.
\begin{table}[h]
\caption{{\label{tab:imed42}}
Coulomb ($V_C$), isovector (MED-$V_C\equiv \beta^{(1)}_{\bf f_{7/2}}$) and isotensor (TED-$V_C\equiv \beta^{(2)}_{\bf f_{7/2}}$)
energies (keV) in $A=42$. $V_C$ calculated in the oscillator basis (ho)} 
\begin{ruledtabular}
\begin{tabular}{ccccc}
&$J=0$&$J=2$&$J=4$&$J=6$\\  
\hline
$V_{C}\equiv V^{\rm ho}_{C\bf{f_{7/2}}}$&\ 81.60&24.60&\-6.40&-11.40\\
$E_J[{^{42}}{\rm Ti}-{^{42}}{\rm Ca}]-V_{C}$&\ \ 5.38&92.55&\ 4.57&-47.95\\
$E_J[{^{42}}{\rm Ti}+{^{42}}{\rm Ca}-2\, ^{42}{\rm Sc}]-V_{C}$&116.76&80.76&\ 2.83&-42.15
\end{tabular} 
\end{ruledtabular}
\end{table}
 
Ironically, this result is obvious from the, long known, $A=42$
spectra~\cite{end90}. Assuming that the observed states are
essentially $f_{7/2}^2$ configurations on top of the $^{40}$Ca core,
these spectra define an interaction in the $f_{7/2}$ subshell.
Therefore, by setting $V_{C}\equiv V^{\rm ho}_{C\bf{f_{7/2}}}$, the
nuclear isovector and isotensor contributions can be extracted. They
are shown in Table~\ref{tab:imed42}, where their centroids ---$\sum_J
(2J+1)V_{\bf{f_{7/2}}}^J/\sum_J (2J+1)$---have been subtracted for
clarity. The assumption of $f_{7/2}^2$ dominance is not warranted,
as---at least--- the $J=0$ and 2 states are known to mix with core
excitations. Therefore a safer procedure consists in replacing the
lowest observed states by the $f_{7/2}^2$ centroids estimated from
spectroscopic factors~\cite{end90}. However, when this is done, no
significant change obtains in Table~\ref{tab:imed42}, whose
indications must therefore be taken very seriously.

A renormalized $V_C$ adapted to the $f_{7/2}^2$ space will remain a
purely $\pi\pi$ force, and therefore {\em the same} for the isovector
and isotensor channels.  Furthermore, it is expected to be reasonably
close to the bare $V_C$ in the first line of Table~\ref{tab:imed42}.
Therefore, upon subtracting this bare $V_C$ from the observed
data---second and third lines---we expect {\em the same, reasonably
  small} numbers. It is obvious that the corresponding numbers are
neither equal nor small. The unavoidable conclusion is that the $A=42$
data indicate that {\em the role of isospin non conserving nuclear
  forces is at least as important as that of the Coulomb potential} in
the observed MED and TED.
 
For the full description of these quantities in $A=46$-51 we rely on
exact, {\em isospin conserving} shell model
calculations~\cite{ANTOINE} with single particle spectrum from
$^{41}$Ca and the KB3G interaction. Very little changes are observed
if the other standard interactions are used (KB3, FPD6, all defined
in~\cite{pov01}). The energy differences are obtained in first order
perturbation theory~\footnote{Diagonalizing instead of taking
  expectation values makes {\em very} little difference.}, as the sum
of expectation values, in which we separate the monopole and multipole
components of the Coulomb field $V_C=V_{Cm}+V_{CM}$ following
Refs.~\cite{duf02,zuk01}:
\begin{eqnarray}
  \label{eq:medj0}
  {\rm MED}_J&=&\Delta_{\rm M}\langle V_{Cm}\rangle_J+\Delta_{\rm M}\langle
  V_{CM}\rangle_J+\Delta_{\rm M}\langle V_{B}\rangle_J,\\
  \label{eq:tedj0}
  {\rm TED}_J&=&\Delta_{\rm T}\langle
  V_{CM}\rangle_J+\Delta_{\rm T}\langle V_{B}\rangle_J.
\end{eqnarray}
The monopole $V_{Cm}$ contains all terms quadratic in scalar products
of Fermion operators $a_i^+\cdot a_j$. The non diagonal contributions
($i\ne j$) lead to isospin mixing that demands second order
perturbation theory. They will be considered here only through their
influence on the radial wavefunctions, i.e., the Thomas Ehrman shift
that depresses the single particle $p_{3/2}$ state in $^{41}$Sc by 200
keV below its analogue in $^{41}$Ca.  The diagonal part ($i=j$)
involves only proton number operators. It contains $E_{C}$ plus a
single particle splitting induced by $V_C$ on the orbits of principal
quantum number $p$ above harmonic oscillator (ho) closed shell
$Z_{cs}$~\footnote{This term can be derived from~\cite{duf02}[Eqs.
  5,10-12]. Hint: the overall coefficient in Eq.~(\ref{eq:epsc}) comes
  from $1.934\times 0.383\times (-0.01)\times 1.522\approx -0.001$.}:
\begin{equation}
  \label{eq:epsc}
\varepsilon_{Cl}=\frac{-0.001\, Z_{cs}^{13/12}\,
  [2l(l+1)-p(p+3/2)]}{A^{1/3}(p+3/2)} {\rm MeV}.  
\end{equation}
The effect of $E_{C}$ is proportional to the difference of (inverse)
radii between a $J$-yrast and the ground state~\cite{50fe}. The total
radii depend on those of the individual orbits, and therefore---to
good approximation---on the {\em average neutron plus proton}
occupancies for each orbit, which we denote by $\langle
m_{k}\rangle_J/2$, with $m_k=z_k+n_k$ (number of neutrons plus number
of protons in orbit $k$). We take averages relying on the near equality
of proton radii in both members of a mirror pair~\cite{duf02}. As it
is reasonable to assume that orbital radii depend only on $l$, and the
$p_{1/2}$ occupancy is always negligible, the whole radial effect will
be taken to depend on the $p_{3/2}$ occupancy. Note that the single
particle contribution from Eq.~(\ref{eq:epsc}) is proportional to the
{\em difference of proton and neutron} occupancies. It is important in
$A=41$, but typically ten times smaller than the radial effect in
$A=$47-51, so we neglect it and end up with $\Delta_M\langle
V_{Cm}\rangle_J=a_m\langle m_{p_{3/2}}\rangle_J/2$.  The value of
$a_m$ can be estimated by {\em adding} to the observed shift the
single particle splitting~(\ref{eq:epsc}) that depresses the $l=3$
orbits with respect to the $l=1$ ones (by 125 keV at $A=40$,
$Z_{cs}=20$). Then, $a_m\approx .200+.125=0.325$ MeV.

{\em In the
isotensor case the $m_{p_{3/2}}$ contributions cancel out.}

The multipole contribution $\Delta\langle V_{CM}\rangle_J$ is
calculated using oscillator Coulomb matrix elements in the $pf$
shell. 
\begin{figure}[h]
  \begin{center}
    \leavevmode 
    \epsfig{file=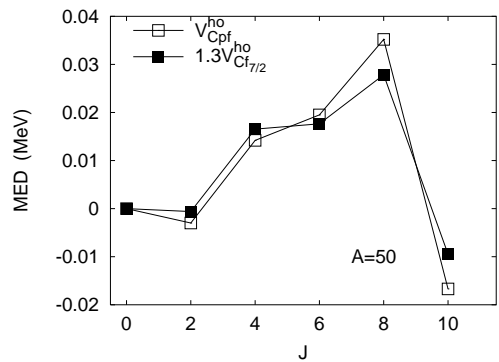,width=7cm}
    \caption{Example of MED renormalization in $A=50$}
    \label{fig:tuttir}
  \end{center}
\end{figure}
The only direct information on $V_B$ comes from
Table~\ref{tab:imed42}. To make use of it we must explore the
possibility of specifying an interaction acting in the full $pf$
shell, {\em solely in terms of $f_{7/2}$ matrix elements}. The idea
turns out to be quite viable using the multiplicative
prescription\footnote{This form proves sufficient for our present
  needs, but it is case-dependent and one can find examples that
  demand a term in $a\, \langle m_{p_{3/2}}\rangle_J$.}
\begin{equation}
  \label{eq:horen}
\Delta\langle V_{C\bf{pf}}^{ho}\rangle_J=b\, \Delta\langle
V_{C\bf{f_{7/2}}}^{ho}\rangle_J,   
\end{equation}
as illustrated for the Coulomb potential in
Fig.~\ref{fig:tuttir}.
 
\begin{figure*}
  \begin{center}
    \leavevmode
    \epsfig{file=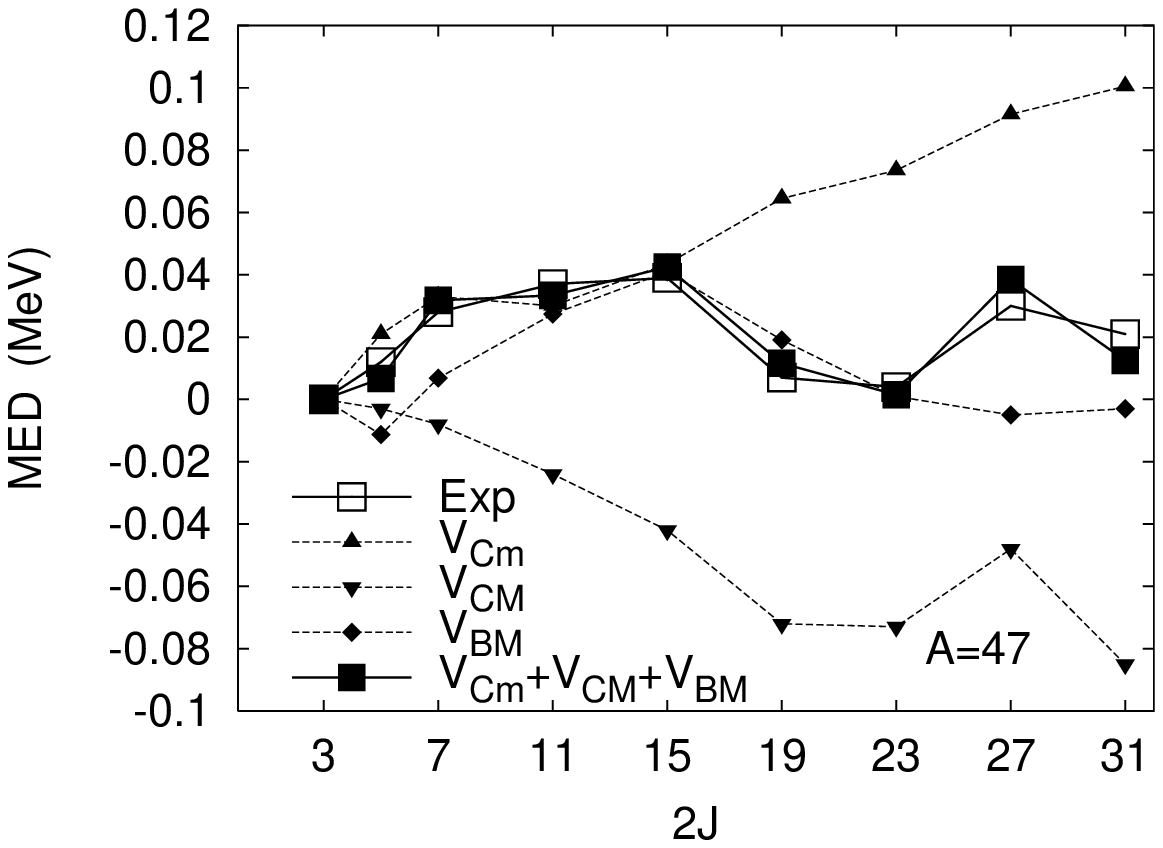,width=7cm}
    \epsfig{file=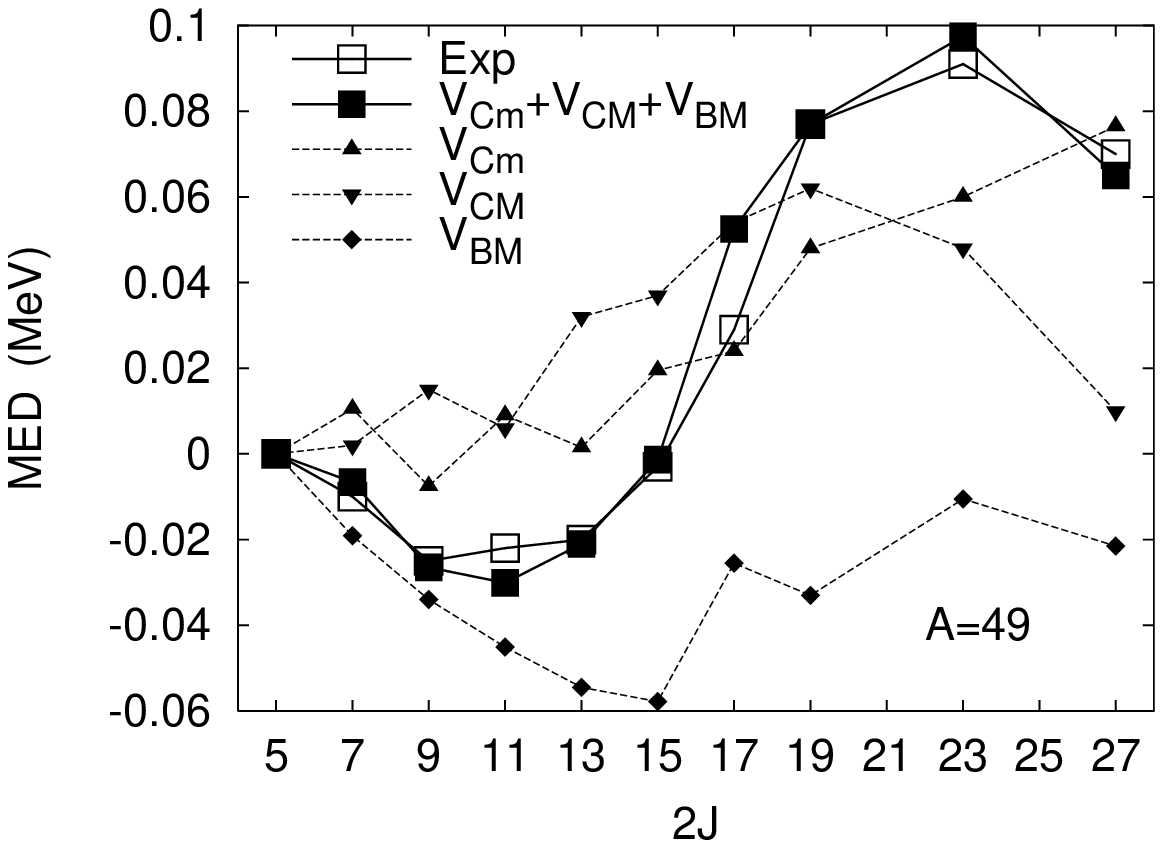,width=7cm}
    \epsfig{file=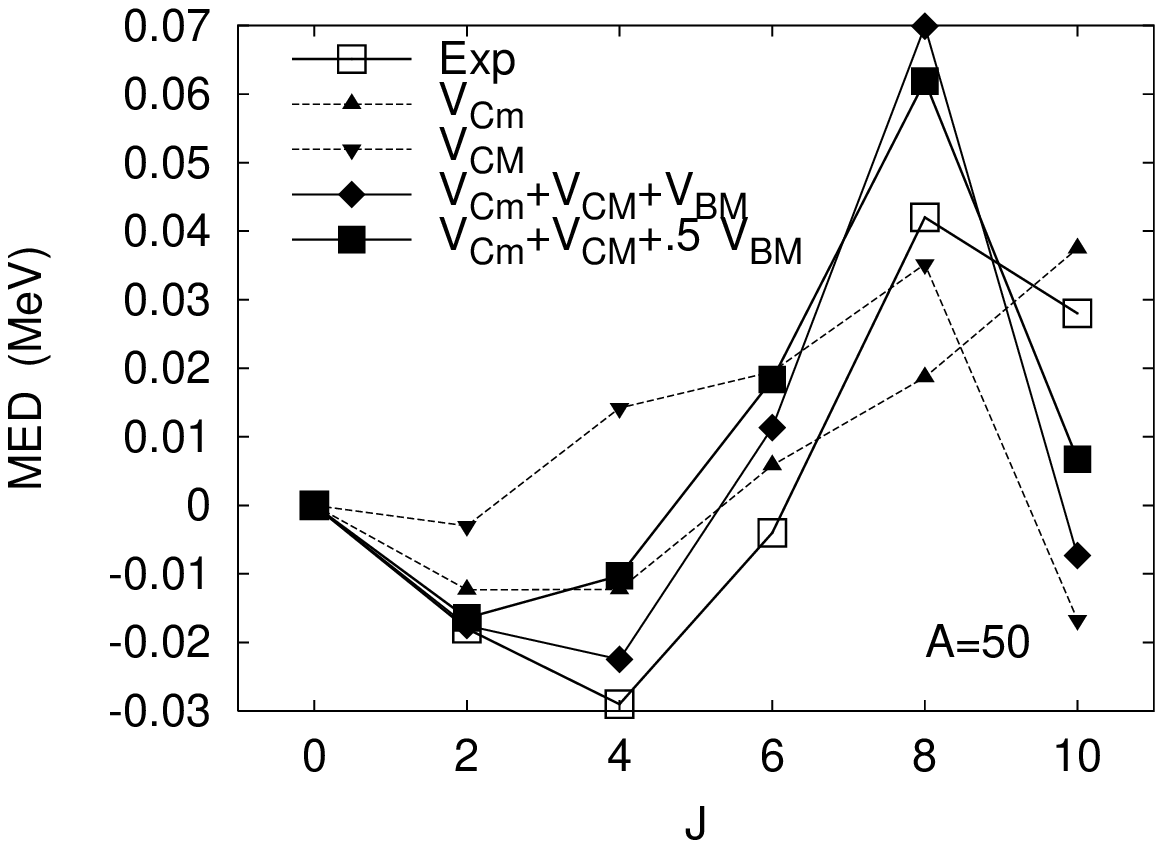,width=7cm}
    \epsfig{file=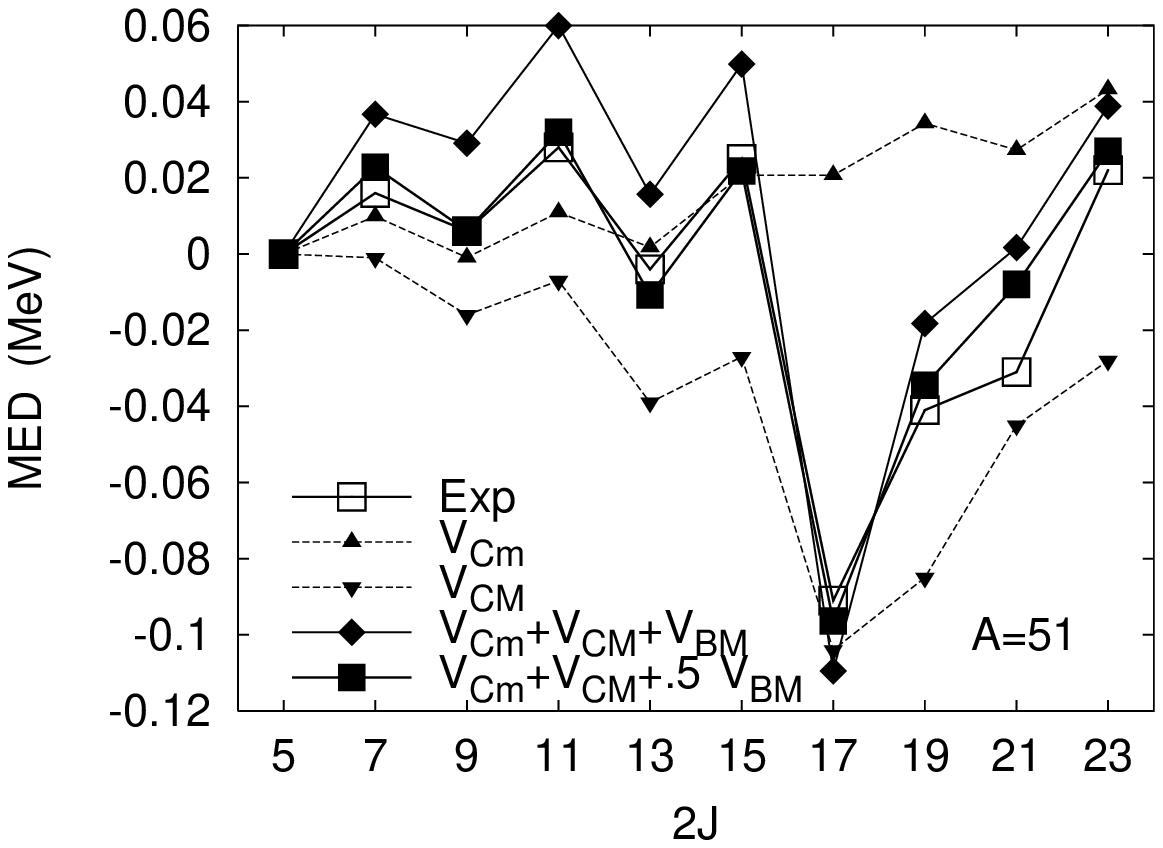,width=7cm}
    \caption {Experimental~\protect\cite{49,47-49,51,50fe} and
      calculated MED for the pairs $^{47}$V-$^{47}$Cr,
      $^{49}$Cr-$^{49}$Mn, $^{50}$Cr-$^{50}$Fe, and
      $^{51}$Mn-$^{51}$Fe.}
    \label{fig:tutti1}
  \end{center}
\end{figure*}
The same form efficiently relates the schematic pairing or quadrupole
pairing forces in the $pf$ shell to the $V^{J=0\, {\rm or}\,
  J=2}_{\bf{f_{7/2}}}$ matrix elements. To minimize the number of
parameters, for $V_B$ we retain only the leading term  
suggested by Table~\ref{tab:imed42}, and set 
$\beta_{\bf{pf}}^{(1)}=\beta_1\, V^{J=2}_{\bf{f_{7/2}}}$,
$\beta_{\bf{pf}}^{(2)}=\beta_2\, V^{J=0}_{\bf{f_{7/2}}}$, where
$V^J_{\bf{f_{7/2}}}$ is the matrix element with unit value.
Collecting all the pieces we have

\begin{eqnarray}
  \label{eq:medj}
{\rm MED}_J&=&\frac{1}{2}a_m\langle
m_{p_{3/2}}\rangle_J+\Delta_{\rm M}\langle
V_{C\bf{pf}}^{ho} +\beta_1\, V^{J=2}_{\bf{f_{7/2}}}\rangle_J,\\
  \label{eq:tedj}
{\rm TED}_J&=&\Delta_{\rm T}\langle
V_{C\bf{pf}}^{ho} +\beta_2\, V^{J=0}_{\bf{f_{7/2}}}\rangle_J.
\end{eqnarray}
In Figs.~\ref{fig:tutti1} for the MED, $V_{Cm},\, V_{CM}$, and
$V_{BM}$ stand respectively for the first, second and third terms in
Eq.~(\ref{eq:medj}). The parameters are taken to be round numbers,
$a_m=300$ keV and $\beta_1=\beta_2=100$ keV.
\begin{figure}[h]
  \begin{center}
    \leavevmode 
    \epsfig{file=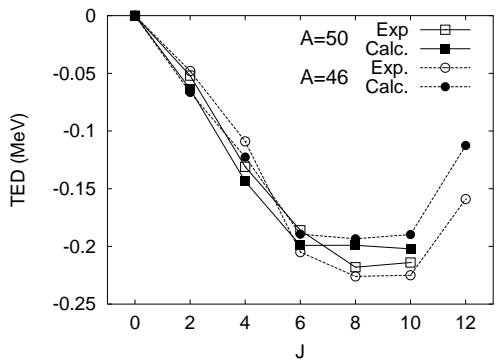,width=7cm}
    \caption{TED for $A=46$ and 50}
    \label{fig:ten}
  \end{center}
\end{figure} 

The reduction of $V_B$, for MED and TED, to a single matrix element is
an oversimplification, but the results are so satisfactory that the
need of extra terms is not felt. The only parameter-free alternative
is to take matrix elements with the weights in Table~\ref{tab:imed42}.
However, this choice is arbitrary because---from the discussion around
Eq.~(\ref{eq:horen}) and Fig.~\ref{fig:tuttir}---we expect a case by
case (even matrix element by matrix element) renormalization.
Nonetheless, though the agreement with experiment becomes less
impressive, it remains acceptable. The conclusion is that the 
leading term in $J=2$ for MED is indeed dominant, and that in $J=0$
for TED very dominant.

The $V_{Cm}$, $V_{CM}$ and $V_{BM}$ contributions, shown
separately in Fig.~\ref{fig:tutti1} for $A=47$ and 49 are quite far
from the observed pattern, which is accurately reproduced 
only after these disparate terms are added.
For $A=50$ and 51 we have replaced the $V_{BM}$ part by a variant of
the full sum in which $\beta_1$ is halved. For $A=50$ the changes are
insignificant, but there is a definite improvement in $A=51$ (remember
again Eq.~(\ref{eq:horen}) and footnote before it). 

It is especially worth noting in Fig.~\ref{fig:tutti1} that the strong
signature effect in the $A=49$ band is erased in the MED by the
out-of-phase $V_{Cm}$ and $V_{CM}$, while the signature staggering is
enhanced in $A=51$. 

The experimental TED patterns in Fig.~\ref{fig:ten} are quite nicely
reproduced by the minimal $\beta_{\bf{pf}}^{(2)}=\beta_2\,
V^{J=0}_{\bf{f_{7/2}}}$ choice. As mentioned, the inclusion of the
$J\ne 0$ terms (third line of Table~\ref{tab:imed42}) makes little
difference, and---interestingly enough---simply {\em ignoring} $V_B$
and {\em doubling} $V_{CM}$ (or the other way round) makes practically
no difference. Which confirms the overwhelming dominance of $J=0$
pairing.

It can be hoped that a rigorous treatment calling upon state of the
art CSB potentials~\cite{mac01} will confirm the role of the $J=2$
pairing term for the isovector MED. The TED behaviour seems far
simpler and our results are consistent with the findings
in~\cite{orm97} for $\beta^{(2)}_{\bf r}$ borne out in~\cite{46cr} for
$A=46$. Therefore, here we may bet on---rather than hope
for---confirmation by the charge independence breaking
potentials~\cite{mac01b}.

The isovector channel raises a difficulty for $A=46$.
In~\cite{46cr,orm97} it was found that $\beta^{(1)}_{\bf r}\approx 0$
using the same functional form as for $\beta^{(2)}_{\bf r}$ with
strong $J=0$ pairing, which does not square with our results. But in
this case our results do not square with experiment either. The scheme
that has been successful in $A=47$, 49, 50 and 51 fails in $A=46$: we
are simply unable to do any better than in~\cite[Fig. 3a)]{46cr}.

\begin{figure}[h]
  \begin{center}
    \leavevmode 
    \epsfig{file=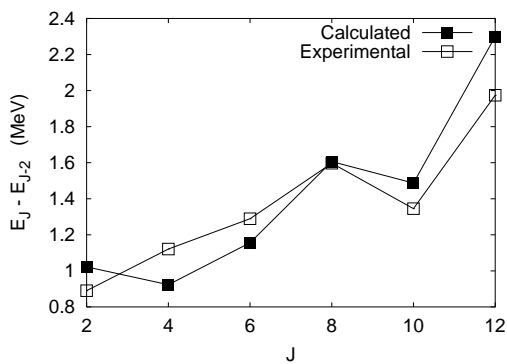,width=7cm}
    \caption{Yrast energy differences in $A=46$}
    \label{fig:46}
  \end{center}
\end{figure}

The trouble is no doubt due to the poor spectroscopy provided by full
$pf$ shell diagonalizations for $A\le 46$, at least when compared with
the very high quality descriptions for the rest of the $f_{7/2}$
nuclei (i.e., $A\le 56$). Fig.~\ref{fig:46} illustrates the point: the
calculated yrast energetics is wrong for the lowest states and, for
the others, far less precise than the corresponding patterns in the
heavier nuclei (see for example~\cite{mar97} for $A=47$ and 49). This
problem extends to transition rates and static moments. It was first
noted and abundantly discussed in Ref.~\cite{pov81} but its
quantitative explanation remains a challenge. This unsatisfactory
situation provides nonetheless a helpful clue: the TED may be
unsensitive to details, but the MED demand accurate wave functions and
could be taken as tests of their quality.

Within the $A=46$ proviso, our results make obvious something that may
seem at first surprising: isospin non conserving potentials play a
role that is at least as important as $V_C$ in explaining the MDE (and
TDE, as found previously in~\cite{46cr}). In this respect, it is worth
noting that {\em direct} evidence for charge symmetry breaking has
been confined, so far, to the {\em very} light systems (basically
$A=2$ and 3)~\cite{mac01}. The mechanism plays an important part in
resolving the Nolen Schiffer anomaly in the MDE, but the effects of
$V_C$ remain much stronger\cite{duf02}. For the MED and TED, $V_C$ is
at most as strong as $V_B$, for which we have shown that substantial
quantitative information can be extracted from the data. To boot, the
MED also provide a view of the evolution of yrast radii.

This work owes much to a stay of AZ at the UAM, made possible by a
scholarship of the BBVA foundation. AP is supported by grant
BFM2000-30 from MCyT (Spain).

\end{document}